\documentclass[11pt]{article}

 \csname
@addtoreset\endcsname{equation}{section}

\def\a{\alpha}  
\def\unA{\underline A}\def\unB{\underline B}
\def\b{\beta}   
\def\c{\gamma} 
\def\C{\Gamma}
\def\d{\delta} 

\def\e{\epsilon} 

\def\f{\phi}

\def\k{\kappa}

\def\L{\Lambda}
\def\m{\mu}
\def\n{\nu}
\def\r{\rho}
\def\s{\sigma}
\def\S{\Sigma}
\def\t{\tau}

\def\O{\Omega}
\def\o{\omega}

\def\cL{{\cal L}}
\def\cF{{\cal F}}

\def\cA{{\cal A}}
\def\cN{{\cal N}}
\def\cR{{\cal R}}
\def\cS{{\cal S}}

\def\cI{{\cal I}}

\def\cA{{\cal A}}

\def\yb{{\bar y}}

\def\unA{\underline{A}}
\def\unB{\underline{B}}

\let\la=\label
\let\bm=\bibitem
\def\nn{\nonumber}
\newcommand{\eq}[1]{(\ref{#1})}

\newcommand{\w}[1]{\\[0.#1cm]}
\def\eqs#1#2{(\ref{#1}-\ref{#2})}

\def\be{\begin{equation}}
\def\ee{\end{equation}}
\def\bea{\begin{eqnarray}}
\def\eea{\end{eqnarray}}
\def\ba{\begin{array}}
\def\ea{\end{array}}

\def\mx#1#2#3#4{\left#1\begin{array}{#2} #3 \end{array}\right#4}

\def\ft#1#2{{\textstyle{{\scriptstyle #1}
\over {\scriptstyle #2}}}}

\def\ket#1{|#1\rangle}

\begin{document}

\hfill{CTP-TAMU-34/01}

\hfill{UU-01-08}

\hfill{hep-th/0112100}


\vspace{20pt}

\begin{center}


{\Large \bf 7D Bosonic Higher Spin Gauge Theory: \\[10pt]Symmetry
Algebra and Linearized Constraints}
\\ \vspace{20pt}
{\sf E. Sezgin}\\
{\it\small Center for Theoretical Physics, Texas A\&M University,
College Station, TX 77843, USA}\\ \vspace{15pt}
{\sf P. Sundell}\\
{\it\small Department for Theoretical Physics, Uppsala University,
Box 803, S-75108, Uppsala, Sweden}


\vspace{30pt}

{\bf Abstract}

\end{center}

We construct the minimal bosonic higher spin extension of the 7D
AdS algebra $SO(6,2)$, which we call $hs(8^*)$. The generators,
which have spin $s=1,3,5,...$, are realized as monomials in
Grassmann even spinor oscillators. Irreducibility, in the form of
tracelessness, is achieved by modding out an infinite dimensional
ideal containing the traces. In this a key role is played by the
tree bilinear traces which form an $SU(2)_K$ algebra. We show that
gauging of $hs(8^*)$ yields a spectrum of physical fields with
spin $s=0,2,4,...$ which make up a UIR of $hs(8^*)$ isomorphic to
the symmetric tensor product of two 6D scalar doubletons. The
scalar doubleton is the unique $SU(2)_K$ invariant 6D doubleton.
The spin $s\geq 2$ sector comes from an $hs(8^*)$-valued one-form
which also contains the auxiliary gauge fields required for
writing the curvature constraints in covariant form. The physical
spin $s=0$ field arises in a separate zero-form in a
`quasi-adjoint' representation of $hs(8^*)$. This zero-form also
contains the spin $s\geq 2$ Weyl tensors, i.e. the curvatures
which are non-vanishing on-shell. We suggest that the $hs(8^*)$
gauge theory describes the minimal bosonic, massless truncation of
M theory on AdS$_7\times S^4$ in an unbroken phase where the
holographic dual is given by $N$ free $(2,0)$ tensor multiplets
for large $N$.

\pagebreak

\setcounter{page}{1}


\section{Introduction}


A higher spin (HS) gauge theory is a general covariant field
theory with an additional infinite set of local symmetries. The
full set of symmetries is based on a rigid HS algebra, which
typically extends the spacetime (super) AdS group. The first step
in the construction of a HS gauge theory is therefore to identify
this algebra and find its representations. Extensive results along
these lines already exist in the literature for $D\leq 5$
\cite{4dv1,kv2,us3,us4,vr2,us1,5dv2,us2}. In this paper we
construct the minimal bosonic HS algebra in $D=7$ and its
linearized massless field equations. Altogether these results show
that the HS theories in diverse dimensions have many features in
common, which points to some underlying universal principle.

As will be discussed in more detail in \cite{holo}, HS gauge
theories in diverse dimensions that include {\it massive} higher
spin fields, are the anti-holographic duals of free conformal
field theories in the limit of large number of free fields
\cite{ed,su2,us1,edseminar}. HS field equations do not seem to
admit any truncation to ordinary (super)gravity, because lower
spin fields appear as sources of HS fields. This is consistent
with the fact that the stress-energy tensor of a free CFT does not
form a closed OPE algebra, as has been illustrated in the case of
the 3d supersingleton \cite{tanii} (though the conserved charges
of course form the finite dimensional conformal group which
closes). Thus, if one wishes to give an anti-holographic
description of a renormalization group flow in the vicinity of a
(free) CFT with conserved HS currents one should use HS gauge
theory instead of ordinary (super)gravity. In particular, the HS
theory discussed in this paper is the minimal bosonic truncation
of a supersymmetric HS theory which is \cite{holo} is proposed to
contain solutions describing the flow from the strongly coupled
$A_{N-1}$ fixed point \cite{Rev} to the free theory of $N$ $(2,0)$
tensors.

Our construction of $hs(8^*)$ is given in terms of a basic
Grassmann even oscillator $y_\a$ which is a Dirac spinor of
$SO(6,1)$. The generators of $hs(8^*)$ are traceless monomials in
the oscillators. The trace is taken with the symmetric charge
conjugation matrix. Importantly, the tracelessness condition is
imposed as a coset condition, namely $hs(8^*)={\cal L}/{\cal I}$
where ${\cal L}$ is a certain Lie algebra which contains both
traceless generators and generators with non-zero trace, and
${\cal I}\subset{\cal L}$ is an ideal containing all the
generators with non-zero trace.

In fact, the above construction is similar to the one of $hs(2,2)$
in $D=5$ \cite{us1,5dv2}, which is the minimal bosonic HS
extension of $SO(4,2)$. The ideal ${\cal I}$ occurring in that
case contains the trace $K\sim \yb y$. The spectrum of the 5D
theory is given by the product of two scalar doubletons, which
have vanishing $K$ charge. This yields massless fields in $D=5$
with spin $s=0,2,4,...$. The spin $s=0$ field and the on-shell
curvatures (Weyl tensors) are collectively described in HS theory
by a master field $\Phi$ which is a zero-form. As a consequence of
modding out ${\cal I}$ the master field $\Phi$ is dressed up by
`$K^2$-expansions' which are crucial for getting the correct AdS
lowest weights \cite{us1,5dv2}.

In $D=7$, the ideal ${\cal I}$ is generated by the traces,
$K_3\sim \yb y$, $K_-\sim y y$ and $K_+\sim \yb\yb$, which form an
$SU(2)_K$ algebra (note that the charge conjugation matrix is
symmetric in $D=7$ whereas it is antisymmetric in $D=5$). Massless
fields in AdS$_7$ arise in tensor products of two 6d conformal
tensors known as doubletons \cite{7dgun}\footnote{In the $\cN=2$
superextension of the 7D theory the superdoubleton squares belong
to an isolated series \cite{FS1,FS2}. On the other hand, the
massless fields of the $D=5$, $\cN=4$ higher spin theory
\cite{us2}, which arise in the square of a certain superdoubleton,
saturate the bound of a continuous series \cite{FS1,FS2}. Thus the
5D theory can be Higgsed in a continuous fashion \cite{holo}.}.
The spectrum of the theory is given by the product of two scalar
doubletons. The scalar doubleton is the only one with vanishing
$SU(2)_K$ charge (in the $(2,0)$ superextension of the theory this
scalar is one of the five scalars in the tensor multiplet). As in
five dimensions, this yields massless fields of spin
$s=0,2,4,...$. We will show that the gauging of $hs(8^*)$ in $D=7$
yields a matching set of gauge fields. In particular, we verify
that for each spin $s$ the algebra generators are in one-to-one
correspondence with the previously known set of physical and
auxiliary spin $s$ gauge fields given in terms of a particular set
of spin $s$ Lorentz tensors \cite{vhd2,m5}. Moreover, modding out
the ideal $\cI$ the 7D master scalar field $\Phi$ is dressed up
with $K^2$-expansions, where now $K^2=K_I K_I$. It is gratifying
that these expansions indeed lead to the correct critical
mass-terms in the linearized equations for $\phi$ and the Weyl
tensors such that the AdS energies assume the correct massless
values.

This paper is organized as follows. In Section 2 we describe the
doubleton representations of the AdS$_7$ group and compute
explicitly the massless 7D field content from the product of two
scalar doubletons. In Section 3 we construct the bosonic HS
algebra $hs(8^*)$. In Section 4 we gauge $hs(8^*)$ by introducing
a master gauge field in the adjoint representation and a master
scalar field in a quasi-adjoint representation of $hs(8^*)$, and
in particular we compute the $K^2$-dressing. In Section 5 we write
the linearized curvature constraints and show that they yield the
correct spectrum of massless fields. Our results are summarized
and further discussed in Section 6. Appendix A contains the
details of the calculation of the mass terms occurring in the
linearized field equations. Appendix B contains the harmonic
analysis on AdS$_7 \times S^4$ needed in the computation of the
AdS energies.


\section{$SO(6,2)$ Representation Theory}


Physical representations of $SO(6,2)$ can be obtained from a set
of Grassmann even oscillator $y_\a$ ($\a=1,..,8$), which is a
$SO(6,1)$ Dirac spinor, obeying the oscillator algebra\footnote{We
use mostly positive signature and Dirac matrices $(\C^a)_\a{}^\b$
obeying $\{\C^a,\C^b\}=2\eta^{ab}$ ($a=0,1,2,3,4,5,7$). The charge
conjugation matrix $C_{\a\b}$ is symmetric and real and
$(\C^aC)_{\a\b}$ are anti-symmetric. The Dirac conjugate of $y_\a$
is defined by $\yb_\a=(y^\dagger i\C^0C)_\a$.}

\be y_\a\star \bar{y}_\b-\bar{y}_\b\star y_\a=2C_{\a\b}\
,\la{osc} \ee
where $\star$ denotes the product of the oscillator algebra. We
also define a Weyl ordered product as follows:

\be y_\a\bar{y}_\b=\bar{y}_\b y_a=y_\a \star \bar{y}_\b-C_{\a\b}\
. \la{weyl} \ee
The Weyl ordered product extends in a straightforward fashion to
contraction rules between arbitrary polynomials of the oscillators
similar to those used in expanding products of Dirac matrices. For
example,

\be (\yb_\a y_\b )\star (\yb_\c y_\d)= \yb_\a \yb_\c y_\b y_\d+
C_{\b\c} \yb_\a y_\d -C_{\a\d} \yb_\c y_\b - C_{\b\c}C_{\a\d}\
.\ee
The $SO(6,2)$ generators are given by ($A=a,8$):

\be M_{AB}=\ft12 \yb \star \S_{AB} y=\ft12 \yb \S_{AB} y\ ,\quad
\S_{a8}={i\over 2}\C_a\ ,\quad \S_{ab}={i\over 2}\C_{ab}\ ,\ee \be
[M_{AB},M_{CD}]=i\eta_{BC}M_{AD}+\mbox{$3$ more}\ ,\quad
(M_{AB})^\dagger=M_{AB}\ .\ee
The spin group $Spin(6,2)$ is a non-compact version of $SO(8;{\bf
C})$, which we denote by $SO(8^*)$. The 7D spin group $Spin(6,1)$
is a subgroup of $SO(8^*)$.

The maximal compact subgroup $SO(6,2)$ is given by $L^0 =
SO(6)\times U(1)_E$ where $SO(6)$ is the group of spatial
rotations and $U(1)_E$ is generated by the the AdS energy
$E=M_{08}$. There is a three-grading $SO(6,2)=L^- \oplus L^0
\oplus L^+$, where $L^{\pm}$ contain non-compact energy-lowering
and energy-raising operators $M^\pm$ such that $[E,M^\pm]=\pm
M^\pm$. The three-grading also requires $[L^+,L^+]=[L^-,L^-]=0$,
$[L^+,L^-]=L^0$ and $[L^\pm,L^0]=L^\pm$. The reality conditions
are $(L^0)^\dagger=L^0$ and $(L^+)^\dagger=L^-$. Positive energy
representations of $SO(6,2)$ are weight spaces
$D(E_0;n_1,n_2,n_3)$ labeled by the lowest energy $E_0$ and
$SO(6)$ highest weight labels $(n_1,n_2,n_3)$ obeying $n_1\geq n_2
\geq |n_3|$. In case of integer spin these determine an $SO(6)$
Young tableaux with $n_1$ boxes in the first row, $n_2$ boxes in
the second row and $|n_3|$ boxes in the third row. If $n_3\neq 0$
the Young tableaux can be self-dual or anti-self-dual, which
corresponds to $n_3>0$ and $n_3<0$, respectively. The $SO(6)$
Young tableaux can be converted into an $SU(4)$ Young tableaux by
contracting it with $SO(6)$ Dirac matrices. A column with $N$
boxes ($N=1,2,3$) stacked on top of each other is contracted with
the rank $N$ Dirac matrix. Their symmetry properties are as
follows: $(\s^R)_{ij}$ ($R=1,...,6$, $i,j=1,...,4$) is
anti-symmetric, $(\s^{RST})_{ij}$ and $(\s^{RST})^{ij}$ are
symmetric and with definite self-duality properties, and
$(\s^{RS})_i{}^j$ belongs to $4\times \bar{4}$. In the case of
half-integer spin all three highest weight labels are
half-integers determining a $\s$-traceless tensor-spinor whose
tensor structure is determined as above by the highest weight
labels $(n_1-\ft12,n_2-\ft12,n_3-\e(n_3)\ft12)$, where $\e(n_3)$
is the sign of $n_3$. For example, $(1,0,0)$ denotes the (real)
$6$-plet, $(1,1,\pm 1)$ denotes the self-dual and anti-self-dual
(complex) $10$-plets, $(\ft12,\ft12,\pm\ft12)$ denotes the chiral
$SU(4)$ spinors and $(\ft32,\ft12,\pm\ft12)$ has the $(1,0,0)$
tensor structure and denotes a chiral and $\s$-traceless
vector-spinor.

Our next aim is to describe the UIRs of $SO(6,2)$ which arise in
the oscillator construction \cite{7dgun}. A convenient choice of
Dirac matrices is

\be C=\mx{(}{cc}{0&1\\1&0}{)}\ ,\quad
\C^0=\mx{(}{cc}{i&0\\0&-i}{)}\ .\ee
The Dirac spinor oscillator $y_\a$ obeying \eq{osc} then split
into two sets of $U(4)$-covariant oscillators ($i=1,...,4$):

\be y_\a=\sqrt{2}(a^i,b_j)\ ,\quad (a_i)^\dagger = a^i\ ,\quad
(b_i)^\dagger=b^i\ ,\la{decomp}\ee\be [a_i,a^j]_*=\d_i^j\ ,\quad
[b_i,b^j]_*=\d_i^j\ .\ee
The $U(4)$-covariant oscillators have a unitary representation in
the Fock space built on the vacuum state $\ket{0}$ defined by

\be a_i\ket{0}=b_i\ket{0}=0\ .\ee
The identification \eq{decomp} implies that the Fock space is a
unitary representation space of $SO(6,2)$. Other unitary
representations of $SO(6,2)$ can then be obtained by considering
tensor products of several copies of the Fock space.

The Fock space and its tensor products decompose into physical
UIR's of $SO(6,2)$. The three-grading $L^- \oplus L^0 \oplus L^+$
takes the following form in the (single) Fock space
representation:

\bea L^+&:& L^{ij}=a^{[i}\star b^{j]}\ ,\\L^0&:& L^i_j=a^i\star
a_j+ b^i\star b_j+4\d^i_j\ ,\\ L^-&:& L_{ij}=a_{[i}\star b_{j]}\
.\eea
The energy operator, which is the trace-part of $L^i_j$, is
bounded from below and contains a constant contribution when it is
written in normal ordered form:

\be E={i\over 4}\yb\star \C^0 y={1\over 4}y^\dagger\star y=
\ft12(a^i\star a_i+b^i \star b_i)+2=\ft12(N_a+N_b)+2\ .\ee
Since $L^0=U(4)\simeq SU(4)\times U(1)_E$ where $SU(4)$ is the
diagonal sum of $SU(4)_a$ and $SU(4)_b$, it is possible to form
lowest weight states (lws) carrying the same $L^0$ weight by
exchanging $a^i$ and $b^i$ oscillators. To be more precise, the
lws carry an extra label of the $SU(2)_K$ group under which
$(a^i,b^i)$, $i=1,...,4$ transform as doublets. In fact, the
$SU(2)_K$ is generated by:

\bea K_+&=&{i\over 4} \yb\yb=-ib^i a_i\ ,\\ K_-&=&{i\over
4}yy=ia^i b_i\ ,\\ K_3&=& {1\over 4}\yb y={1\over 4}\yb\star
y+2=\ft12(N_b-N_a)\ .\la{sp1}\eea
For computational purposes it is convenient to write ($I=1,2,3$)

\be K_I={1\over 8} y^{\unA}(\s_I)_{\unA\unB}y^{\unB}\ ,\quad
y_{\unA}=y^{\unB}\O_{{\unB}{\unA}}=(y_\a,\yb_\a)\ ,\la{ki1}\ee
where $(\s_I)_{\unA\unB}$ are symmetric Pauli matrices tensored
with the $SO(8^*)$ charge conjugation matrix and $y_A$ obeys

\be y_{\unA}\star y_{\unB}=y_{\unA} y_{\unB}+\O_{\unA\unB}\ ,\quad
\O_{\unA\unB}=\mx{(}{cc}{0&C_{\a\b}\\-C_{\a\b}&0}{)}\ .
\la{ki2}\ee
By construction $[SO(6,2),SU(2)_K]_\star=0$.  Thus the Fock space
and its tensor products decompose into $(2j+1)$-plets of identical
$SO(6,2)$ weight spaces, which we denote by
$D^{(j)}(E_0;n_1,n_2,n_3)$.

The Fock space $\cF$ of a single set of oscillators decomposes
into $SO(6,2)\times SU(2)_K$ weight spaces, known as doubletons,
as follows \cite{7dgun}:

\be \cF=\sum_{s=0,\ft12,1,\ft32,...} D^{(s)}(E_0=s+2;s,s,\pm s)\
.\la{fock}\ee
The lws with $K_3$-eigenvalue $m$ is given by

\be a^{(i_1}\cdots a^{i_{2s-2m}}b^{i_{2s-2m+1}}\cdots
b^{i_{2s})}\ket{0}\ .\ee
The doubletons describe off-shell conformal tensors in six
dimensions\footnote{The compact $SO(6)$ labels are related to the
non-compact $SO(5,1)$ labels, i.e. the six-dimensional spin
labels, and the energy $E_0$ to the six-dimensional scaling
dimension by a non-unitary rotation \cite{7dgunnc}.} with spin $s$
and scaling dimension $E_0$. In particular, the $SU(2)_K$-singlet
describes a scalar field with scaling dimension $2$.

The $P$-fold tensor product $\cF^{\otimes P}$ of the oscillator
Fock space $\cF$ can be described by adding a flavor-index
($r,s=1,...,N$):

\be [a_i(r),a^j(s)]=\d_{rs}\d_i^j\ ,\quad
[b_i(r),b^j(s)]=\d_{rs}\d_i^j\ .\ee
The representation of $SO(6,2)\times SU(2)_K$ on the tensor
product is given by:

\be L_{ij}=\sum_r L_{ij}(r)\ ,\quad L^i{}_j=\sum_r L^i{}_j(r)\
,\quad L^{ij}=\sum_r L^{ij}(r)\ ,\ee \be \quad K_I=\sum_r K_I(r)\
.\ee
In particular, the total energy operator is given by:

\be E=2P+{1\over 2}\sum_r \left[N_a(r)+N_b(r)\right]\ .\ee
Thus the tensor product decomposes into weight spaces with lowest
energy $E_0\geq s+2P$ (the lws with $E_0>s+2P$ are obtained by
anti-symmetrizing oscillators carrying different flavor indices).
In particular, $P=2$ yields massless fields in AdS$_7$
\cite{7dgun}\footnote{These fields also satisfy the masslessness
criteria defined in \cite{FF}. In \cite{FF} certain UIRs of
$SO(6,2)$ that cannot be obtained by multiplying any number of 6d
doubletons. However, the unitarity bounds of $Osp(8^*|4)$ seem to
exclude this possibility \cite{Mi2,FS3}. }.


\section{The Massless Spectrum of The Theory}


As discussed in the Introduction, the minimal bosonic 7D HS theory
is conjectured to be the anti-holographic dual of the 6d theory of
$N$ free scalar fields $\varphi^i$ , i.e. $N$ copies of the scalar
doubleton, in the limit of large $N$ \cite{edseminar,holo}. The
composite operators of this theory couple to (non-normalizable)
bulk modes of the HS theory. In particular, the bilinear operators
are the spin $s=0$ operator $\varphi^i\varphi^i$ and a set of
conserved, symmetric and traceless tensors of spin $s=2,4,6,...$
\cite{vcurrents}. These tensors are in one-to-one correspondence
with the massless representations in the symmetric tensor product
of two 6d scalar doubletons. The anti-symmetric tensor product
contains massless representations which correspond to descendants.
The massless spectrum of our 7D theory is therefore given by:

\be \cS=[D(2;0,0,0)\otimes D(2;0,0,0)]_S\ .\la{spectrum}\ee

In order to decompose $D(2;0,0,0)\otimes D(2;0,0,0)$ under
$SO(6,2)$ we compute the lws with energy $E_0=4+s$ for
$s=0,1,2,...$. To this end, we expand a general state
$\ket{\psi}\in D(2;0,0,0)\otimes D(2;0,0,0)$ with that energy as
follows:

\be \ket{\psi}=\sum_{\m=0}^s\ket{\psi^{(\m)}}=\sum_{\m=0}^s
\psi^{(\m)}{}_{i_1j_1,\dots, i_\m j_\m;k_1l_1,\dots,
k_{s-\m}l_{s-\m}} L^{i_1j_1}(1)\cdots L^{i_\m j_\m}(1)L^{k_1
l_1}(2)\cdots L^{k_{s-\m} l_{s-\m}}(2)\ket{0}\ .\ee
The quantities $L^{i_1j_1}(1)\cdots L^{i_\m j_\m}(1)$ and $L^{k_1
l_1}(2)\cdots L^{k_{s-\m} l_{s-\m}}(2)$ are irreducible under
$SU(4)$. Their $SO(6)$ highest weight labels are $(\m,0,0)$ and
$(s-\m,0,0)$, respectively. Thus the $SU(4)$ tensors $\psi^{(0)}$
and $\psi^{(n)}$ are irreducible, and given by the $SU(4)$ Young
tableaux with two rows of length $s$. This $SU(4)$ irrep, which we
shall denote by $R_s$, has spin given by $s$ and can be converted
to a real, symmetric rank $s$ $SO(6)$ tensor by contracting it
with $s$ (anti-symmetric) $SO(6)$ Dirac matrices. Its $SO(6)$
highest weight labels are $(s,0,0)$. The remaining $\psi^{(\m)}$,
$0<\m<s$, are reducible and decompose into $R_s$ plus a set of
various other irreps, $\{R^{(\m)}\}$ say (where each irrep occur
once and only once), which we write as

\be
\psi^{(\m)}=\psi^{(\m)}(R_s)+\sum_{R^{(\m)}}\psi^{(\m)}(R^{(\m)})\
,\quad 0<\m<s\ .\ee
The state $\ket{\psi}$ is a lws provided that

\be L_{ij}\ket{\psi}\equiv (L_{ij}(1)+L_{ij}(2))\ket{\psi}\equiv
\sum_{\n=0}^{s-1}\ket{\chi^{(\n)}}=0\ ,\ee
where $\ket{\chi^{(\n)}}$ contains $\n$ factors of $L^+(1)$ and
$s-1-\n$ factors of $L^+(2)$. Thus

\be \ket{\chi^{(\n)}}=0\ ,\quad \n=0,\dots,s-1\ .\ee
The state $\ket{\chi^{(\n)}}$ is a linear combination of
contributions from $\psi^{(\n)}$ and $\psi^{(\n+1)}$. From
$\ket{\chi^{(0)}}=0$ and the fact that the contributions from the
various irreps are linearly independent it follows that

\be \psi^{(1)}(R_s)=-s^2 \psi^{(0)}\ ,\quad \psi^{(1)}(R^{(1)})=0\
.\ee
From $\ket{\chi^{(1)}}=0$ it then follows that

\be \psi^{(2)}(R_s)=-{(s-1)^2\over 4}\psi^{(1)}(R_s)={s \choose
2}^2\psi^{(0)}\ ,\quad \psi^{(2)}(R^{(2)})=0\ .\ee
Iterating this procedure we find that there is precisely one lws
with energy $E_0=4+s$, which belongs to the $SU(4)$ irrep $R_s$
described above. Collecting the above results, we find the
following explicit expression for this state:

\be \ket{E_0;s,0,0}=\sum_{\m=0}^s (-1)^\m{ s\choose \m}^2
L^{i_1j_1}(1)\cdots L^{i_\m j_\m}(1)L^{i_{\m+1}j_{\m+1}}(2)\cdots
L^{i_s j_s}(2)\ket{0}\ ,\la{lws}\ee
where separate symmetrization of the $i$ and $j$ indices is
assumed. The lws with even spin belong to the symmetric tensor
product and those with odd spin to the anti-symmetric product. In
summary:

\bea \left[D(2;0,0,0)\otimes
D(2;0,0,0)\right]_{\rm S}&=&\sum_{s=0,2,4,...} D(E_0=s+4;s,0,0)\ ,\la{cs}\\
\left[D(2;0,0,0)\otimes D(2;0,0,0)\right]_{\rm
A}&=&\sum_{s=1,3,5,...} D(E_0=s+4;s,0,0)\ .\eea
We have thus computed the massless spectrum of our minimal bosonic
HS theory, which hence consists of massless states with spin
$s=0,2,4,...$ and vanishing $SU(2)_K$ charge.


\section{The Higher Spin Algebra $hs(8^*)$}


Our next task is to determine the HS symmetry algebra, $hs(8^*)$.
From the boundary point of view, $hs(8^*)$ is the algebra of
charges of the set of conserved currents built from the $s\geq 2$
symmetric traceless tensors \cite{vcurrents}. These rigid
symmetries of the free CFT induce local symmetries in the
anti-holographic bulk theory, including general covariance.
However, instead of determining $hs(8^*)$ from the current algebra
we construct it directly in terms of the oscillators. We do this
by making use of the properties of the spectrum derived in the
previous section, and the knowledge of which gauge fields are
required for writing the covariant constraints for a massless
field of given spin $s\geq 2$ in a linearization around AdS$_7$
\cite{vhd2,m5}. In fact, the `canonical' set of spin $s$ gauge
fields which contains physical as well as auxiliary fields is in
one-to-one correspondence with the above mentioned set of
conserved currents of that spin \cite{vcurrents}.

Thus, the methodology we adopt is to impose constraints on general
oscillator expansions. Essentially we impose three types of
constraints: 1) we project out all monomials except those which
are of degree $4\ell+2$, where $\ell=0,1,2,...$ is a level index;
2) we impose neutrality under $SU(2)_K$; and 3) we mod out an
ideal which contains all the traces\footnote{The last step leads
to an algebra which cannot be represented on doubletons that carry
non-zero $SU(2)_K$ charge.}.

We begin by taking $\cA$ to be the space of arbitrary polynomials
$f(y,\yb)$ in the oscillators, which is an associative algebra
with a $\star$-product defined by the Weyl ordering \eq{weyl}. An
element of $\cA$ can therefore be expanded in terms of Weyl
ordered monomials in the basic oscillator $y_\a$ and its Dirac
conjugate $\yb_\a$ with complex coefficients which are
multispinors. We shall use the following normalization convention
(which we give here for a single monomial):

\be f(y,\yb)={1\over m!~n!} ~\yb^{\a_1}\cdots
\yb^{\a_m}y^{\b_1}\cdots y^{\b_n} f_{\a_1\dots\a_m,\b_1\dots\b_n}\
.\ee
We next define a linear anti-automorphism $\tau$ of $\cA$ as
follows:

\be \t(f(y,\yb))= f(iy,i\yb)\ ,\la{tau}\ee \be \t (f_1\star
f_2)=\t(f_2)\star \t(f_1)\ .\la{ai}\ee
A linear anti-automorphisms of an associative algebra can be used
to define a Lie subalgebra. In our case we define the following
Lie subalgebra of $\cA$:

\be \cL=\{P\in\cA: \t(P)=P^\dagger=-P\ ,\quad [K_I,P]_\star=0\}\ ,
\la{tco}\ee
where $K_I$ are the $SU(2)_K$ generators defined in \eq{ki1}. The
Lie bracket of $\cL$ is

\be [P_1,P_2]_\star=P_1\star P_2-P_2\star P_1\ .\ee
The generators of $\cL$ have expansions with multispinor
coefficients which in general have non-zero trace parts. In order
to impose tracelessness we define a new ordering of elements in
$\cA$ which amounts to factoring out the trace parts explicitly
\cite{us1}. An element $P\in\cL$ is thus expanded as:

\be P=\sum_{n=0}^{\infty}P_{(n)}^{I_1\dots I_n}(y,\yb)\star
K_{I_1}\star \cdots \star K_{I_n}\ ,\la{e}\ee
where $P_{(n)}^{I_1\dots I_n}(y,\yb)$ has an expansion in terms of
traceless, Weyl ordered multispinors and the $SU(2)_K$ indices
$I_1\dots I_n$ are symmetric. The conditions on $P_{(n)}^{I_1\dots
I_n}$ imply that

\be K_I\star P_{(n)}^{I_1\dots I_n}=K_I P_{(n)}^{I_1\dots I_n}-{i
n\over 2} \e_{IK}{}_{\phantom{(n)}}^{(I_1}P_{(n)}^{I_2\dots
I_n)K}\ .\ee
The expansion \eq{e} leads to the following decomposition of
$\cL$:

\be \cL=\cL_{(0)}\oplus \cL_{(1)}\oplus
\cL_{(2)}\oplus\cdots=\cL_{(0)}+\cI\ ,\quad \cI=\cL_{(1)}\oplus
\cL_{(2)}\oplus\cdots\ ,\la{ci}\ee
where $\cL_{(n)}$ represents the $n$'th term in \eq{e}. If we were
to change the prescription for ordering the $SU(2)_K$ generators
in \eq{e} this would only affect the multispinors in $\cI$. Thus
the traceless multispinors in $\cL_{(0)}$ are uniquely defined by
\eq{e}.

The space $\cL_{(0)}$ of traceless generators decomposes into
levels labeled by $\ell=0,1,2,...$ consisting of elements of the
form

\be P={1\over
(n!)^2}~P_{\a_1\dots\a_n,\b_1\dots\b_n}\yb^{\a_1}\cdots \yb^{\a_n}
y^{\b_1}\cdots y^{\b_n}\ ,\quad n=2\ell+1\la{yt}\ ,\ee
where the multispinor $P_{\a_1\dots\a_n,\b_1\dots\b_n}$ belongs to
an (irreducible) Young tableaux of $Spin(6,2)\simeq SO(8^*)$ with
two rows of equal length $n$. In order to see this we first note
that from the condition

\be [K_3,P]_\star=0\ \ee
it follows that $P$ must have an equal number of $y$ and $\yb$
oscillators. The conditions

\be [K_\pm,P]_\star =0\ee
then set to zero the $SO(8^*)$ irreps in $P$ described by Young
tableaux which have more boxes in the first row than in the second
row.

The reality condition $P^\dagger=-P$ implies that the real
dimension of \eq{yt} equals that of the corresponding $SO(8)$
Young tableaux. The Young tableaux is reducible under the 7D spin
group $Spin(6,1)\subset SO(8^*)$. It decomposes into irreps
labeled by (tensorial) $SO(6,1)$ Young tableaux with two rows of
which the first one has $n$ boxes and the second one $m$ boxes,
where $0\leq m \leq n$. The counting works out simply because
$SO(8^*)$ decomposes under $Spin(6,1)$ in the same way as
$SO(6,2)$ decomposes under $SO(6,1)$. The conversion from the
$SO(8^*)$ irrep to the set of $SO(6,1)$ irreps can be explicitly
done by contracting the $SO(8^*)$ Young tableaux with $m$ second
rank Dirac matrices $(\C^{ab})_{\a\b}$ and $n$ first rank Dirac
matrices $(\C^c)_{\a\b}$, where the $SO(6,1)$ indices belong to
the $SO(6,1)$ Young tableaux. We express the resulting generators
as:

\bea {1\over (n!)^2}
P^{(m,n-m)}_{\a_1\dots\a_n,\b_1\dots\b_n}\yb^{\a_1}\cdots
\yb^{\a_n}y^{\b_1}\cdots y^{\b_n}
&=&P^{a_1}_{b_1}{}^{\dots}_{\dots}{}^{ a_m}_{b_m}{}^{ c_1\dots
c_{n-m}}M^{b_1}_{a_1}\cdots M^{b_m}_{a_m}P_{c_1}\cdots
P_{c_{n-m}}\ ,\nn\\  n=2\ell+1\ ,\quad 0\leq m \leq n\
.&&\la{no}\eea
Upon gauging, these generators lead to the canonical set of gauge
fields with spin $s=2\ell+2$ ($\ell=0,1,2,...$) as discussed
earlier. In particular, the zeroth level consists of the $SO(6,2)$
generators $P_a$ and $M_{ab}$ corresponding to the gravitational
gauge fields.

The space $\cI$ defined in \eq{ci} forms an ideal:

\be [\cL,\cI]_*\subset \cI\ .\ee
Moreover, there is a redundancy such that content of $\cL_{(0)}$
is reproduced in the $SU(2)_K$ trace part of $\cL_{(n)}$ for
$n=2,4,...$. For example, the generators $P\delta^{IJ}\star
K_I\star K_J$ give rise to one copy of $\cL_{(0)}$, and so on. We
are therefore led to defining the HS algebra by

\be hs(8^*)=\cL/\cI\ .\ee
The elements of $hs(8^*)$ are thus equivalence classes $[P]$ of
elements in $\cL$ defined by

\be [P]=\{P'\in \cL\ |\ P'-P\in \cI\}\ .\la{hs}\ee
The Lie bracket of $[P_1]$ and $[P_2]$ is given by

\be [[P_1],[P_2]]=[[P_1,P_2]_*]\ .\ee

In order to examine the representation theory of the HS algebra
$hs(8^*)$ we observe the following grading of $hs(8^*)$:

\bea hs(8^*)&=&\sum_{n\in \mathbf Z} L^{(n)}\ ,\qquad
[E,L^{(n)}]_\star
= n L^{(n)}\ ,\\
{}[L^{(n)},L^{(m)}]_\star &=& L^{(m+n)}\ ,\qquad
(L^{(n)})^{\dagger}=L^{(-n)}\ .\eea
Moreover, from \eq{no} it follows that $hs(8^*)$ is a subalgebra
of the universal enveloping algebra $Env(SO(6,2))$:

\be hs(8^*)=Env(SO(6,2))/\cR\ ,\la{env}\ee
where $\cR$ is an ideal generated by various polynomials in
$Env(SO(6,2))$ which vanish by Fierz identities that arise when
the single oscillator realization of $SO(6,2)$ is used. By a
choice of ordering we can thus write

\be L^{(n)}=\left\{ (L^+)^p\star(L^0)^{\star ~q}\star(L^-)^r:
p-r=n;\ p+q+r=2\ell+1,\ \ell=0,1,2,... \right\}/\cR\ ,\la{d000}\ee
where $L^-\oplus L^0\oplus L^+$ is the three-grading of $SO(6,2)$,
$\ell$ is the level index and we are using the non-commutative
$\star$-product. From \eq{d000} it follows that $L^{(0)}\sim
L^0+(L^0)^3 +\cdots $ when acting on an $SO(6,2)$ lowest weight
state. A physical representation of $hs(8^*)$ is thus a lowest
weight representation $\widehat{D}(E_0;m_1,m_2,m_3)$ based on an
$SO(6,2)$ lowest weight state $\ket{E_0;m_1,m_2,m_3}$ obeying the
additional conditions

\be L^{(-n)}\ket{\O}=0\ ,\quad n=1,2,3,...\ .\la{hslws}\ee
Since the single oscillator vacuum state $\ket{0}$ obeys
\eq{hslws}, the scalar doubleton, which is the fundamental
$SO(6,2)$ UIR, is also the fundamental UIR of $hs(8^*)$:
$$\widehat{D}(2;0,0,0)=D(2;0,0,0)\ .$$ Tensor products of the scalar
doubleton also form $hs(8^*)$ representations. If $P\in hs(8^*)$
then the representation of $P$ on the $N$-fold tensor product
$(D(2;0,0,0))^{\otimes N}$ is given by $P= \sum_{r=1}^N P_r$,
where $P_r$ acts on the $r$th factor in the tensor product. Thus,
in particular, the massless spectrum $\cS$ given in \eq{spectrum}
is an $hs(8^*)$ representation. The $SO(6,2)$ lowest weight state
$\ket{4;0,0,0}=\ket{0}\in \cS$ is an $hs(8^*)$ lowest weight.
There are no other $hs(8^*)$ lws in $\cS$. To see this we first
note that any $hs(8^*)$ lws must also be an $SO(6,2)$ lws. It thus
suffices to show that for each lws in \eq{lws} with $s=2,4,6,...$
there exists at least one energy-lowering operator in $hs(8^*)$
which does not annihilate this state. To exhibit such an operator
we first write \eq{lws} in the following schematic form
($s=2,4,6,...$):

\be \ket{s+2;s,0,0}= \left((L^+_1)^s+(L^+_1)^{s-1}L^+_2+\cdots +
(L^+_2)^s\right)\ket{0}\ .\ee
Acting on $\ket{s+2;s,0,0}$ with the element $L^+\star
(L^-)^{s}\in L^{(-s+1)}$ (the level of this state is given by
$\ell=s/2$) yields

\be \left[ L^+_{\phantom{1}}\star (L^-)^s\right]\ket{s+2;s,0,0}=
\left[L^+_1\star (L^-_1)^s+L^+_2\star (L^-_2)^s\right]\ket{\psi}
=(L^+_1+L^+_2)\ket{0}\neq 0\ .\ee
It follows that the massless spectrum $\cS$ defined in
\eq{spectrum} is an irreducible $hs(8^*)$ multiplet:

\be \cS=\widehat{D}(4;0,0,0)\ .\la{hsm}\ee


\section{Gauging $hs(8^*)$}\la{sec:g}


In order to gauge $hs(8^*)$ we introduce an $hs(8^*)$-valued
one-form $[A]$ and a zero-form $\Phi$ obeying the conditions:

\be \t(A)=A^\dagger=-A\ ,\quad [K_I,A]_\star=0\ ,\la{irr1}\ee \be
\t(\Phi)=\Phi^{\dagger}=\pi(\Phi)\ ,\quad K_I\star \Phi=\Phi\star
K_I=0\la{phi}\ .\la{irr2}\ee
Here $\t$ is the anti-automorphism defined in \eqs{tau}{ai} and
$\pi$ is an automorphism acting on $SU(2)_K$-invariant elements
$f\in\cA$ as follows:

\be \pi(f^{(m,n)})=(-1)^n f^{(m,n)}\ ,\quad \pi(f_1\star
f_2)=\pi(f_1)\star \pi(f_2)\ ,\la{pi}\ee
where $m$ and $n$ are related to the $SO(6,1)$ highest weight
labels as:

\be m_1=m+n\ ,\quad m_2=m\ ,\quad m_3=0\ .\ee
Here we have used the fact that any $SU(2)_K$ invariant element
$f\in\cA$ can be expanded in terms of such $SO(6,1)$ irreps as
explained in the previous section; see the analysis following
\eq{yt}. In order to show that $\pi$ is an automorphism, we make
use of the fact that if $f_1, f_2\in\cA$ are $SU(2)_K$ invariant
then also $f_1\star f_2$ is $SU(2)_K$ invariant. Thus $f_1\star
f_2$ can also be expanded in terms of $SO(6,1)$ Young tableaux
with $m_3=0$. Using the basic rules for $SO(6,1)$ tensor products
one can then verify that $\pi$ is an automorphism.

The conditions on $\Phi$ defines a representation of $hs(8^*)$
which we call quasi-adjoint. It is essential to introduce this
representation to accommodate the physical scalar field and the
spin $s\geq 2$ Weyl tensors as well as all the derivatives of
these fields. The curvature and covariant derivative are defined
by

\bea F_{[A]}&=&[F_A]=[dA+A\star A]\ ,\quad
D_{[A]}\Phi=d\Phi+A\star\Phi-\Phi\star\pi(A)\ ,\\
\d_{[\e]}[A]&=& [d\e+[A,\e]_\star]\ ,\quad \d_{[\e]}\Phi=
\e\star\Phi-\Phi\star\pi(\e)\ .\la{gt}\eea
The conditions on $\Phi$ implies that $D_{[A]}\Phi$ and
$\d_{[\e]}\Phi$ are also quasi-adjoint elements and that they do
not depend on the choice of $A$ and $\e$. The role of the $\pi$
automorphism is to distinguish between the translations-like and
rotations-like generators in the HS algebra; for
example\footnote{For $D=4,5$, the map $\pi$ can also be defined as
a linear transformation of the oscillators. This does not seem to
be  possible in $D=7$, since $(\C_a C)_{\a\b}$ and
$(\C_{ab}C)_{\a\b}$ have the same symmetry. It is worth noting
that there is an alternative way of implementing the $\pi$ map by
extending the oscillator algebra $\cA$ with an inner Kleinian
operator $\k=(-1)^{2K_3}$. The associative algebra
$\tilde{\cA}=\cA\oplus (\k\cA)$ has the $\pi$ map
$\pi(f(y,\yb;\k)=f(y,\yb;-\k)$, and the AdS generators can be
taken to be $\tilde{P}_a=\k P_a$ and $\tilde{M}_{ab}=M_{ab}$. More
generally, $hs(8^*)$ is generated by $\kappa^{m+\s}
P^{(m,2\ell+1-m)}$ ($m=0,\dots,2\ell+1$) for $\s=1$. Taking
$\tilde{\Phi}$ to obey \eq{irr2} and imposing the curvature
constraints \eqs{c1}{c2}, which are now expansions in $\k$, one
finds that $\s=0$ gives rise to an entire set of auxiliary gauge
fields. Thus the physical field content remains the same in this
framework. }:

\be \pi(P_a)=-P_a\ ,\quad \pi(M_{ab})=M_{ab}\ .\ee
Hence, if $\O$ denotes an $SO(3,2)$ valued connection, we find
that $D_\O\Phi= \nabla\Phi+dx^\m\{P_\m,\Phi\}_{\star}$, where
$\nabla$ is the Lorentz covariant derivative. As will be shown in
in the next section, this means that the whole one-form $D_\O\Phi$
can be constrained without having the consequences of setting
$\Phi$ to constant.

We next solve the conditions on $\Phi$. By subtracting the two
$K_I$-conditions in \eq{phi} we obtain $[K_I,\Phi]_\star =0$. Thus
we can choose to expand $\Phi$ as follows:

\be \Phi=\sum_{n=0}^\infty \Phi_{(n)} f_{(n)}=\sum_{n=0}^\infty
 {1\over n!} \Phi^{I_1\dots I_n}_{(n)}(y,\yb)
K_{I_1}\cdots K_{I_n} f_{(n)}(K^2)\ ,\la{phi1}\ee
where $\Phi^{I_1\dots I_n}_{(n)}(y,\yb)$ consists of traceless
multispinors which are also traceless in their internal $SU(2)_K$
indices, i.e.

\bea K_I\star \Phi_{(n)}^{I_1\dots I_n}&=&K_I \Phi_{(n)}^{I_1\dots
I_n}-{i n\over 2} \e_{IK}{}^{(I_1}\Phi_{(n)}{}^{I_2\dots I_n)K}\
,\\ \delta_{JK} \Phi_{(n)}^{JKI_1\dots I_{n-2}}&=&0\ ,\quad n>1\
.\eea
The quantities $f_{(n)}(K^2)$ are analytical functions of
$K^2=K_IK_I$, which we determine below. Note that had we chosen to
use $\star$-products of $SU(2)_K$ generators in \eq{phi1}, instead
of classical products, the $K_I$-condition would only admit the
trivial solution $\Phi=0$.

It thus remains to impose $K_I\star\Phi=0$. After some algebra,
where it is convenient to use \eq{ki1} and \eq{ki2}, we find

\be K_I\star \left( \Phi_{(n)} f_{(n)}\right) = K_I\Phi_{(n)}
D[f_{(n)}] -4\left(\mid \Phi_{(n)}\mid +4\right) {\partial
\Phi_{(n)}\over
\partial K_I} f_{(n)}\ ,\ee
where $D$ is the second order differential operator

\be D[f_{(n)}(z)]=f_{(n)}(z)-{1\over 8}(7+\mid \Phi_{(n)}
\mid)f'_{(n)}(z)- {1\over 4}zf''_{(n)}(z)\ ,\ee
and $\mid\Phi_{(n)}\mid$ denotes the number of irreducible spinor
indices in the multispinor defining $\Phi_{(n)}$:

\be y^{\unA}\partial_{\unA}\Phi^{I_1\dots I_n}(y,\yb)=\mid
\Phi_{(n)} \mid \Phi^{I_1\dots I_n}(y,\yb)\ .\ee
From $K_I\star \Phi=0$ it follows that\footnote{ A solution to
$K_I\star \Phi=\Phi\star K_I=0$ can also be constructed formally
as $\Phi=\Pi\star \Phi$, where $\Pi$ is the projector of the
single oscillator Fock space \eq{fock} onto the spin $s=0$
doubleton, which is the $SU(2)_K$-singlet.}

\be f_{(n)}(K^2)=\mx{\{}{ll}{F(\mid \Phi_{(0)} \mid;K^2)&n=0\\
0&n>0}{.}\ee
where the function $F(w;z)$ is given by

\be F(w;z)= \sum_{n=0}^{\infty} {(4z)^n\over
n!}{\C(\ft12(w+7))\over \C(\ft12(w+7+2n))}\ .\la{F}\ee

From $[K_I,\Phi_{(0)}]_\star=0$, and following the analysis below
\eq{yt}, it follows that the $y$ and $\yb$ expansion of
$\Phi_{(0)}(y,\yb)$ gives rise to traceless multispinors
$\Phi_{(0)\a_1\dots\a_s,\b_1\dots \b_s}$ that belong to spin $s$
$SO(8^*)$ Young tableaux that have two rows of equal length $s$.
Taking into account also the condition $\t(\Phi)=\pi(\Phi)$ we
find (from now on we drop the subscript $(0)$)

\bea \Phi(y,\yb)&=&\sum_{\tiny\ba{c}
m=0,2,4,...\\n=0,1,2,...\dots\ea}
\Phi^{(m,n;0)}(y,\yb)F(2s;K^2)\\&=& \sum_{\tiny\ba{c}
m=0,2,4,...\\n=0,1,2,...\dots\ea} \sum_{k=0}^\infty
\Phi^{(m,n;k)}(y,\yb)\ ,\la{fexp}\eea
where the $y$ and $\yb$ expansion of $\Phi^{(m,n;0)}(y,\yb)$
yields a traceless multispinor $\Phi^{(m,n;0)}_{\a_1\dots
\a_{s},\b_1\dots \b_{s}}$ which is equivalent to a spin $s$
$SO(6,1)$ Young tableaux with $s=m+n$ boxes in the first row and
$m$ boxes in the second row. The multispinor
$\Phi^{(m,n;k)}_{\a_1\dots \a_{s+k},\b_1\dots \b_{s+k}}$
($k=1,2,3,...$) which by definition contain $k$ traces times an
$SO(6,1)$ Young tableaux with $s=m+n$ boxes in the first row and
$m$ boxes in the second row, is given by \eq{F} and \eq{fexp}. For
example

\bea \Phi^{(m,n;1)}_{\a_1\dots \a_{s+1},\b_1\dots\b_{s+1}}&=&0\ ,\nn\\
\Phi^{(m,n;2)}_{\a_1\dots \a_{s+2},\b_1\dots\b_{s+2}}&=&
{(s+2)^2(s+1)^2\over 2(7+2s)}\Phi^{(m,n;0)}_{\a_1\dots
\a_{s},\b_1\dots\b_{s}}\nn\\&&\times
(C_{\a_{s+1}\b_{s+1}}C_{\a_{s+2}\b_{s+2}}-
C_{\a_{s+2}\a_{s+2}}C_{\b_{s+1}\b_{s+2}})\
,\phantom{\b_{s+1}\b_{s+2}}\la{trpart}\eea
where separate symmetrization in the $\a$ and $\b$ indices is
assumed.

The spin $s=0$ sector of $\Phi$ is a single real scalar field
$\f$ with the following $K^2$-dressing:

\be \f(1+{8\over 7}K^2+\cdots)\ ,\la{kd}\ee
where the coefficient $8/7$ is read off from \eq{F}. The spin
$s=1$, i.e. $|\Phi|=2$, sector contains an $SO(6,1)$ vector
$\f_a$. In the spin $s=2$ sector, i.e. for $\mid \Phi \mid=4$ we
find a symmetric traceless tensor $\f_{ab}$ and a traceless tensor
$C_{ab,cd}=C_{cd,ab}$ obeying $C_{[ab,c]d}=0$. These tensors match
the on-shell second order derivatives of a scalar and a graviton,
respectively. This pattern extends to higher spins such that the
tensorial content of $\Phi$ is isomorphic to the derivatives of a
spin $s=0$ field and a set of spin $s=2,4,6,...$ Weyl tensors
(obeying Klein-Gordon equations and Bianchi identities for $s\geq
2$).

\section{The Linearized Curvature Constraints}

In analogy with the 5D case \cite{us1}, we propose the following
linearized curvature constraints ($s=2,4,6,...$):

\bea F^{\rm lin}_{\a_1\dots \a_{s-1},\b_1\dots
\b_{s-1}}&=&e^a\wedge e^b (\C_{ab})^{\c\d}
\Phi^{(s,0;0)}_{\c\a_1\dots \a_n,\d\b_1\dots \b_n}\
,\la{c1}\\[10pt]
D_{\O}\Phi&=&0\ ,\la{c2}\eea
where $\O$ is the AdS background

\be \O=i(P^a e_a+\ft12 M^{ab} \o_{ab}) \ ,\quad d\O+\O\wedge\star
\O=0\ ,\ee
and $F^{\rm lin}$ is the linearized curvature:

\be F^{\rm lin}=dA+\O\star A+ A \star \O\ .\la{flin}\ee
These constraints are invariant under the linearized form of the
gauge transformations \eq{gt}:

\be \d_{\e}A=d\e+[\O,\e]_{\star}\ ,\quad
\d_{\e}\Phi=\e\star\Phi-\Phi\star\pi(\e)\ .\la{lingt}\ee

Consistency of the constraints requires integrability, i.e.
$d^2=0$. The integrability of the scalar constraint \eq{c2}
follows immediately from the flatness of $\O$. The integrability
of the curvature constraint \eq{c1} is equivalent to the Bianchi
identity $dF^{\rm lin}+[\O,F^{\rm lin}]_\star=0$, which takes the
following form in components:

\bea &&\nabla^{\phantom{l}}_{[\m}F^{\rm lin}_{\n\r],\a_1\dots
\a_{s-1},\b_1\dots \b_{s-1}}\nn\\&&+\ft{(s-1)^2}2\left(
(\C^{\phantom{l}}_{[\m})_{\a_1}{}^{\c}F^{\rm
lin}_{\n\r]\,\a_2\dots \a_{s-1}\c,\b_1\dots \b_{s-1}}-
(\C^{\phantom{l}}_{[\m})_{\b_1}{}^{\c}F^{\rm
lin}_{\n\r]\,\a_1\dots \a_{s-1},\b_2\dots \b_{s-1}\c}\right) =0\
,\la{bi}\eea
where separate symmetrization in $\a$ and $\b$ indices is assumed
and $\nabla_\m$ is the Lorentz covariant derivative In order to
verify this we use the component form of the scalar constraint
\eq{c2} which reads:

\be \nabla_\m\Phi_{\a_1\dots\a_n,\b\dots \b_n}+ \ft12
(\C_\m)^{\c\d} \Phi_{\c\a_1\dots\a_n,\d\b\dots \b_n} +\ft{n^2}2
(\C_\m)_{\a_1\b_1}\Phi_{\c\a_2\dots\a_n,\d\b_2\dots \b_n} =0\
,\la{c3}\ee
where again separate symmetrization in $\a$ and $\b$ indices is
assumed. Inserting the curvature constraint \eq{c1} into the
Bianchi identity \eq{bi} and making use of \eq{c3} we find that
\eq{bi} decomposes into two irreducible parts. In the notation
defined in \eq{no} these are given by an $(s,0)$ part and an
$(s-1,1)$ part (by construction these equations have no trace
parts). These are satisfied due to the following Fierz identities
($s\geq 2$):

\bea
(\C_{a[b})^{\a\b}(\C_{cd]})^{\c\d}\Phi^{(s,0;0)}_{\a\c\e_1\dots\e_{s-2},
\b\d\f_1\dots\f_{s-2}}&=&0\ ,\\
(\C_{[a})^{\a\b}(\C_{bc]})^{\c\d}\Phi^{(s-1,1;0)}_{\a\c\e_1\dots\e_{s-2},\b\d
\f_1\dots\f_{s-2}}&=&0\ .\eea
In order to verify these it is important to use the symmetries of
$\Phi^{(s,0;0)}$ and $\Phi^{(s-1,1;0)}$ as implied by the
properties of their $SO(6,1)$ Young tableaux. To implement these
properties it is convenient to use expansions similar to \eq{tb}.

The scalar constraint \eq{c2} implies that the independent fields
in $\Phi$ are given by $\Phi^{(s,0;0)}$ ($s=0,2,4,...$) and that
the remaining components are derivatives of these fields;
schematically $ \Phi^{(s,n;0)} \sim \nabla^n\Phi^{(s,0;0)}$. The
constraint also implies the following Klein-Gordon equations (the
derivation of the mass-term for arbitrary $s$ is given in Appendix
A):

\be (\nabla^2-m^2)\Phi^{(s,0;0)}=0\ ,\quad m^2=-8-2s\
.\la{crit}\ee
Using the harmonic analysis in Appendix B, we find the following
lowest energy of $\Phi^{(s,0;0)}$:

\be E_0=s+4\ .\la{es}\ee
As discussed in Section 2, this is the correct value for a
massless spin $s$ field. Thus the independent field content of the
quasi-adjoint representation $\Phi$ is isomorphic to the spectrum
$\cS$ in \eq{spectrum}. We remark that the spectrum $\cS$ is a
massless $hs(8^*)$ multiplet. The global $hs(8^*)$ transformations
on $\Phi$ are realized in terms of gauge transformations \eq{gt}
with rigid parameters $\e$ obeying the Killing equation

\be D_\O\e=0\ .\la{ke}\ee

The $K^2$-expansion found in the previous section play a crucial
role in obtaining the correct critical mass-value in \eq{crit}.
Let us demonstrate this in the case of spin $s=0$. We examine the
leading equations in the spin $s=0$ sector of \eq{c2}:

\bea \partial_\m\f&=&-{1\over 2}(\C_\m)^{\a\b}\Phi_{\a,\b}\ ,\la{se1}\\
\nabla_\m\phi_{\a,\b}&=&-{1\over 2}(\C_\m)^{\c\d}\Phi_{\a\c,\b\d}
-{1\over 2}(\C_\m)_{\a\b}\f\ .\eea
It follows that

\be \nabla^\m\partial_\m\f={1\over
4}\left((\C^\m)^{\a\c}(\C_\m)^{\b\d}
\Phi_{\a\b,\c\d}+(\C^\m)^{\a\b}(\C_\m)_{\a\b}\f\right)\ .\ee
The second term on the right hand side contributes to $m^2$ by
$-14$. The multispinor in the first term can be decomposed into
$SO(6,1)$ irreps as follows:

\be \Phi_{\a\b,\c\d}=\Phi^{(2,0;0)}_{\a\b,\c\d}+
\Phi^{(1,1;0)}_{\a\b,\c\d}+ \Phi^{(0,2;0)}_{\a\b,\c\d}+
\Phi^{(0,0;2)}_{\a\b,\c\d}\ .\ee
The contributions to $m^2$ from the first three terms vanish, due
to the Fierz identity

\be (\C^\m)^{\a\c}(\C_\m)^{\b\d}\Phi^{(m,n;0)}_{\a\b,\c\d}=0\ ,
\quad m+n=2\ . \ee
The $SO(6,1)$ singlet $\Phi^{(0,0;2)}_{\a\b,\c\d}$ yields a
non-zero contribution to $m^2$ given by

\be {1\over 4}(\C^\m)^{\a\c}(\C_\m)^{\b\d}~{2\over 7}{1\over 4}
\left(2C_{\a\c}C_{\b\d}+2C_{\a\d}C_{\b\c}-4C_{\a\b}C_{\c\d}
\right)=6\ .\ee
As a result we find that the spin $s=0$ field obeys

\be (\nabla^\m\partial_\m-m^2)\f=0\ ,\quad m^2=-14+6=-8\ ,\ee
which leads to the lowest weight energy $E_0=4$.

The curvature constraint \eq{c1}, when written in $SO(6,1)$
tensorial basis, is of the canonical form \cite{vhd2,m5}. Thus the
gauge fields consist of ($s=2,4,6,...$)

\bea \mbox{generalized vielbeins}&:&
A_{\m\,\a_1\dots\a_{s-1},\b_1\dots \b_{s-1}}^{(0,s-1)}\la{gv}\\
\mbox{auxiliary gauge fields}&:&
A_{\m\,\a_1\dots\a_{s-1},\b_1\dots \b_{s-1}}^{(1,s-2)},
\dots\dots, A_{\m\,\a_1\dots\a_{s-1},\b_1\dots
\b_{s-1}}^{(s-1,0)}\phantom{x\qquad\qquad x}\la{ag}\eea
The auxiliary gauge fields can be solved for in terms of
derivatives of the generalized vielbeins. The linearized field
equations, which are second order equations, are obtained by
solving for $A^{(1,s-2)}_\m$ from $F^{(0,s-1)}_{\m\n}=0$ and
substituting into a certain projection of the constraint on
$F^{(1,s-2)}_{\m\n}$. Upon fixing a Lorentz-like gauge, one finds
that the physical fields carry the representations $D(s+4;s,0,0)$.

In summary, as a result of the constraints \eq{c1} and \eq{c2} the
independent set of fields in the theory are

\be \left\{~\phi,\  A_{\m}{}^{a},\  A_{\m}{}^{abc},\
A_{\m}{}^{abcde},\  \dots\dots \phantom{{1\over 1}}\right\}\
.\la{phf}\ee
which obey second order field equations for massless fields
carrying the representations in the spectrum $\cS$ given in
\eq{spectrum} and \eq{cs}.


\section{Conclusions}


We have constructed a minimal bosonic HS extension of the AdS$_7$
group $SO(6,2)$, which we call $hs(8^*)$, consisting of generators
of spin $s=1,3,5,...$. We have realized this symmetry in a 7D HS
gauge theory with massless fields of spin $s=0,2,4,...$, which are
given in \eq{phf}. The $s\geq 2$ fields are contained in a
$hs(8^*)$ valued one-form and the spin $s=0$ field, the Weyl
tensors and their derivatives are assembled in a zero-form which
transforms in a quasi adjoint representation of $hs(8^*)$. The
spectrum of physical fields form a UIR of $hs(8^*)$ which is
isomorphic to the symmetric product of two scalar doubletons. The
gauge fields \eqs{gv}{ag} and the spin $s=0$ field correspond to
the full set of conserved bilinear currents \cite{vcurrents} and
the quadratic `mass' term $\varphi^i\varphi^i$, respectively, of a
6d theory of free scalar fields $\varphi^i$ ($i=1,...,N$) which
carry the scalar doubleton representation.

The internal $SU(2)_K$ algebra given in \eq{ki1} plays a key role
for embedding $hs(8^*)$ and its quasi-adjoint representation in an
associative oscillator algebra. Irreducibility is achieved by
imposing the $SU(2)_K$ invariance conditions on the master fields
as in \eqs{irr1}{irr2}, choosing the ordering prescriptions \eq{e}
and \eq{phi1} and modding out the ideal $\cI$ defined in \eq{ci}.
The $SU(2)_K$ invariance implies that a 6d realization of
$hs(8^*)$ in terms of doubletons must necessarily be given in
terms of the scalar doubleton (which is part of the tensor
multiplet in the case of $(2,0)$ supersymmetry).

The structure of the above 7D HS theory suggests that it describes
the massless sector of a $hs(8^*)$ gauge theory which includes
massive fields and whose holographic dual is the 6d scalar field
theory in the limit of large $N$ \cite{edseminar,holo}. The scalar
doubleton theory is the minimal bosonic truncation of the theory
of $N$ free $(2,0)$ tensor multiplets. We expect that the latter
theory has an anti-holographic dual for large $N$ which is a 7D
gauge theory based on a superextension of $hs(8^*)$ with $\cN=2$
supersymmetry. Once we have computed the interactions in the 7D
theory, these ideas can be tested explicitly by comparing an 6d
$n$-point functions to the corresponding amplitude of the bulk
theory. Since both sides are weakly coupled when $N$ is large
\cite{edseminar,holo} this provides an explicit example of a
directly verifiable AdS/CFT correspondence, similar to the one
proposed in \cite{ed,su2,us1} for $D=5$.

The results obtained this far on HS gauge theories in diverse
dimensions point to underlying universal features. In particular:

\begin{itemize}

\item They are gauge theories of HS algebras that are infinite dimensional
extensions of the finite dimensional AdS group. These algebras are
based on oscillator realizations, or equivalently AdS group
enveloping algebras modded by certain ideals.

\item The massless sector of a HS gauge theory is a UIR of the HS
algebra given by the symmetric tensor product of two
ultra-short multiplets, known as singletons or doubletons, which
decomposes into an infinite tower of AdS supermultiplets where the
first level is the supergravity multiplet.

\item Their massless field content is given by an adjoint gauge
field $A_\m$ and a quasi-adjoint zero-form $\Phi$.

\item Their background independent field equations follow from a
universal set of curvature constraints.

\end{itemize}
These properties and other arguments which will be presented
elsewhere \cite{holo} suggest that HS gauge theories have
holographic duals given by various free, large $N$ conformal field
theories and that finite $N$ corrections are encoded into the bulk
theory in a universal background independent `quantization'
scheme. This would describe an unbroken phase of Type IIB string/M
theory.

Clearly much remains to be done to develop HS gauge theories
further. The supersymmetric extension of the linearized 7D HS
gauge theory presented here, and the linearized 6D HS gauge theory
based on the HS extension of the AdS$_6$ superalgebra $F_4$ should
be straightforward. As for the interactions, they are known fully
in $D=4$ \cite{4dv1} and some cubic couplings have been computed
in the bosonic 5D HS theory \cite{5dv2}. Those in $D=4$ are given
in a closed form but considerable amount of work remains to be
done to exhibit their structure explicitly. The construction of
the full interactions in $D>4$ HS gauge theories also remains an
open problem though we do no expect any fundamental obstacle in
achieving this. In testing ideas of higher spin AdS/CFT
correspondence, it is also important to incorporate the massive HS
multiplets which arise in higher than second order tensor product
of singletons/doubletons \cite{7dgun}. The coupling of massless
and massive HS multiplets is therefore an important open problem.

Another open problem is to understand the role played by the
$U(1)_K$ and $SU(2)_K$ charged higher spin doubletons in $d=4$ and
$d=6$, respectively. They may ultimately be necessary in the
description of an unbroken phase of M theory. In any event, it
seems likely that their inclusion will lead to a HS conformal
field theory in the boundary and corresponding generalized HS
gauge theory in the bulk based on a HS algebra whose maximal
finite dimensional subalgebra is the symplectic extension of the
AdS algebra \cite{5dv1}.

\vspace{10pt}


 \noindent{\Large \bf Acknowledgements}


P.S. is thankful to U. Danielsson and F. Kristiansen for
discussions. This research project has been supported in part by
NSF Grant PHY-0070964.

\pagebreak

\begin{appendix}

\section{Calculation of Mass Terms}

In order to derive the linearized field equation obeyed by the
spin $s$ Weyl tensor
$\Phi^{(s,0;0)}_{\a_1\dots\a_s,\b_1\dots\b_s}$ we start from the
following components of the scalar constraint \eq{c3}:

\bea \nabla_\m\Phi_{\a_1\dots\a_s,\b_1\dots\b_s}&=&-{1\over
2}(\C_\m)^{\c\d}\Phi_{\c\a_1\dots\a_s,\d\b_1\dots\b_s}-{s^2\over
2}(\C_\m)_{\a_1\b_1}\Phi_{\a_2\dots\a_s,\b_2\dots\b_s}\ ,\la{nphi}\\
\nabla_\m\Phi_{\a_1\dots\a_{s+1},\b_1\dots\b_{s+1}}&=&-{1\over
2}(\C_\m)^{\e\f}\Phi_{\e\a_1\dots\a_{s+1},\f\b_1\dots\b_{s+1}}-{(s+1)^2\over
2}(\C_\m)_{\a_1\b_1}\Phi_{\a_2\dots\a_{s+1},\b_2\dots\b_{s+1}}\
.\nn \eea
Here and in the remainder of this section we assume separate
symmetrization of $\a$ and $\b$ indices. Combining the two
equations and restricting to the $(s,0;0)$ sector we find:

\bea \nabla^2\Phi^{(s,0;0)}_{\a(s),\b(s)}&=&
\ft14(\C^\m)^{\c\d}\left[(\C_\m)^{\phantom{()}}_{\c\d}
\Phi^{(s,0;0)}_{\a(s),\b(s)} + s~(\C_\m)^{\phantom{()}}_{\c\b_1}
\Phi^{(s,0;0)}_{\a(s),\d\b(s-1)}\phantom{{1\over2}}\right.\nn\\[9pt]
&&\left.+s~(\C_\m)^{\phantom{()}}_{\a_1\d}
\Phi^{(s,0;0)}_{\c\a(s-1),\b(s)} + \ft14(\C_\m)^{\e\f}
\Phi^{(s,0;2)}_{\c\e\a(s),\d\f\b(s)}\phantom{{1\over2}}\right]\la{m2}\\
&\equiv &m^2\Phi^{(s,0;0)}_{\a(s),\b(s)}\ ,\nn\eea
where $\a(s)=\a_1\dots\a_s$ and idem $\b$. The contribution to
$m^2$ from the first three terms on the right hand side is readily
found to be

\be -14-{7\over 2}s\ .\la{m1}\ee
The calculation of the contribution to $m^2$ from the last term in
\eq{m2} is more elaborate. We need to use \eq{trpart} and break up
the overlapping symmetrizations. In doing so it is convenient to
go over to the $SO(6,1)$ tensorial basis using

\be
\Phi^{(s,0;0)}_{\a(s),\b(s)}=(\C^{a_1}{}_{b_1})_{\a_1\b_1}\cdots
(\C^{a_s}{}_{b_s})_{\a_s\b_s}\Phi^{b_1}_{a_1}{}^{\dots}_{\dots}{}^{
b_s}_{a_s}\ ,\la{tb}\ee
where $\Phi^{a_1}_{b_1}{}^{\dots}_{\dots}{}^{ a_s}_{b_s}$ belongs
to the $SO(6,1)$ Young tableaux with two rows of length $s$.
Making also use of the Fierz identities

\bea
(\C^a{}_{b\m})_{\a_1\b_1}(\C^c{}_d{}^\m)_{\a_2\b_2}\Phi^b_a{}^d_c&=&
-\Phi^{(2,0;0)}_{\a(2),\b(2)}\ ,\\
(\C^a{}_{b\m})_{\a_1\a_2}(\C^c{}_d{}^\m)_{\b_1\b_2}\Phi^b_a{}^d_c&=&
-4\Phi^{(2,0;0)}_{\a(2),\b(2)}\ ,\eea
where the right hand sides are defined as in \eq{tb} and extra
indices on $\Phi$ have been suppressed, we find that the
contribution to $m^2$ from the last term in \eq{m2} is given by

\be {1\over 8(7+2s)}(24s^2+180s+336)=6+{3\over 2}s\ .\la{m22}\ee
Adding the two contributions in \eq{m1} and \eq{m22} we find

\be m^2=-8-2s\ .\la{msquare}\ee

\section{Harmonic Analysis}

To determine the $SO(6,2)$ content of the spectrum, we shall
follow the technique used in \cite{es1} which is based on the
analytic continuation of AdS$_7$ to $S^7$, and consequently the
group $SO(6,2)$ to $SO(8)$. The Casimir eigenvalues for an
$SO(6,2)$ representation $D(E_0;n_1,n_2,n_3)$, where $n_1\geq
n_2\geq |n_3|$ are $SO(6)$ highest weight labels, and an $SO(8)$
representation with highest weight labels $\ell_1\geq \ell_2 \geq
\ell_3\geq |\ell_4|$, are given by

\bea C_2[SO(6,2)] &=& E_0(E_0-6) +n_1(n_1+4)+n_2(n_2+2)+n_3^2
\nn\w2 \ C_2[SO(8)] &=& \ell_1(\ell_1+6) + \ell_2(\ell_2+4) +
\ell_3(\ell_3+2) + \ell_4^2 \ .\la{ha1} \eea
The continuation from AdS$_7$ to $S^7$ requires the
identification:

\be \nabla^2|_{{\rm AdS}_7}\rightarrow -\nabla^2|_{S^7}\ ,\quad
\ell_1=-E_0\ ,\quad \ell_{2,3,4}=n_{1,2,3}\ .\la{ha2}\ee
A tensor $T_{(m_1m_2m_3)}$ on $S^7$ in an irrep $R$ of $SO(7)$
with highest weight labels $m_1\geq m_2\geq m_3\geq 0$ can be
expanded as

\be T_{(m_1m_2m_3)}(x)=\sum_{\tiny
\ba{c}(\ell_1\ell_2\ell_3\ell_4)\\p\ea}
T_p^{(\ell_1\ell_2\ell_3\ell_4)}
D^{(\ell_1\ell_2\ell_3\ell_4)}_{(m_1m_2m_3),p}(L_x^{-1})\
,\la{ha4}\ee
where $L_x$ is a coset representative of a point $x\in S^7$ and
$(\ell_1\ell_2\ell_3\ell_4)$ label all $SO(8)$ representations
satisfying the embedding condition

\be \ell_1\geq m_1\geq \ell_2\geq m_2\geq \ell_3 \geq m_3\geq
|\ell_4|\ .\la{emb}\ee
The Laplacian acting on $T$ yields

\be -\nabla^2|_{S^7}D^{(\ell_1\ell_2\ell_3\ell_4)}_{(m_1m_2m_3),p}
=\left(\phantom{{1\over 2}}\!\!\!
C_2[SO(8)]-C_2[SO(7)]~\right)D^{(\ell_1\ell_2\ell_3\ell_4)}_{(m_1m_2m_3),p}\
,\la{ha5}\ee
where the $SO(7)$ Casimir is given by

\be C_2[SO(7)] = m_1(m_1+5) +m_2(m_2+3) +m_3(m_3+1)\ .\ee
Using the notation introduced in \eq{fexp}, the $SO(7)$ highest
weight labels of $\Phi^{(m,n)}_{\a_1\dots\a_{2s}}$ ($m+n=s$) are

\be m_1=m+n\ ,\quad m_2=m\ ,\quad m_3=0\ .\la{hwrs}\ee
The embedding condition \eq{emb} implies

\be \ell_1\geq s\ ,\quad \ell_2=s\ ,\quad \ell_3= 0,1,...,s\
,\quad \ell_4=0\ .\ee
The on-shell conditions on $\Phi^{(s,0;0)}$, which follow from the
$(s,0;0)$ part of \eq{nphi}, remove the $SO(6,2)$ irreps with
$\ell_3=0,1,...,s-1$. Recalling \eq{msquare} we thus find that the
lowest weight energy of $\Phi^{(s,0;0)}_{\a(s),\b(s)}$ is given by

\be E_0=3+\sqrt{9+s(s+5)+s(s+3)-s(s+4)-8-2s}=4+s\ .\ee
Thus the physical field content in $\Phi^{(s,0;0)}_{\a(s),\b(s)}$
is the massless spin $s$ field carrying the representation
$D(s+4;s,0,0)$.

\end{appendix}

\pagebreak


\end{document}